\documentclass[article,preprint,groupedaddress]{revtex4}
\usepackage{epsfig}
\usepackage{graphicx}
\usepackage{amssymb}

\begin{document}
\title{No nonminimally coupled massless scalar hair for spherically symmetric neutral black holes}
\author{Shahar Hod}
\affiliation{The Ruppin Academic Center, Emeq Hefer 40250, Israel}
\affiliation{ }
\affiliation{The Hadassah Institute, Jerusalem 91010, Israel}
\date{\today}

\begin{abstract}
\ \ \ We provide a remarkably compact proof that spherically
symmetric neutral black holes cannot support static nonminimally
coupled massless scalar fields. The theorem is based on causality
restrictions imposed on the energy-momentum tensor of the fields
near the regular black-hole horizon.

\end{abstract}
\bigskip
\maketitle

\section{Introduction}

The non-linearly coupled Einstein-scalar field equations have
attracted the attention of physicists and mathematicians for more
than five decades. Interestingly, the composed Einstein-scalar
system is characterized by very powerful and elegant no-hair
theorems \cite{Whee,Car,Bekto} which rule out the existence of
asymptotically flat black-hole solutions with regular event horizons
that support various types of scalar (spin-0) matter configurations.

The early no-hair theorems of Chase \cite{Cha} and Bekenstein
\cite{Bek1} have ruled out the existence of regular black holes
supporting static minimally coupled massless scalar field
configurations. The no-hair theorems of Bekenstein \cite{Bek1} and
Teitelboim \cite{Teit} have excluded the existence of black-hole
hair made of minimally coupled massive scalar fields
\cite{Noteit,Hodrc,Herkr}. Later no-hair theorems of Heusler
\cite{Heu} and Bekenstein \cite{Bek2} have ruled out the existence
of neutral black-hole spacetimes supporting static matter
configurations made of minimally coupled scalar fields with positive
semidefinite self-interaction potentials.

The physically interesting regime of scalar fields nonminimally
coupled to gravity has been investigated by several authors. In a
very interesting paper, Mayo and Bekenstein \cite{BekMay} have
proved that spherically symmetric charged black holes cannot support
matter configurations made of charged scalar fields nonminimally
coupled to gravity with generic values of the dimensionless coupling
parameter $\xi$ [The physical parameter $\xi$ quantifies the
nonminimal coupling of the field to gravity, see Eq. (\ref{Eq4})
below]. Intriguingly, the rigorous derivation of a no-hair theorem
for {\it neutral} scalar fields nonminimally coupled to gravity
seems to be a mathematically more challenging task. In particular,
the important no-hair theorems of \cite{BekMay,Saa} can be used to
rule out the existence of spherically symmetric scalar hairy
configurations in the restricted physical regimes $\xi<0$ and
$\xi\geq1/2$ \cite{Notesa}.

The main goal of the present paper is to present a (remarkably
compact) unified no-hair theorem for neutral massless scalar fields
nonminimally coupled to gravity which is valid for {\it generic}
values of the dimensionless coupling parameter $\xi$ (In particular,
below we shall extend the interesting no-scalar hair theorems of
\cite{BekMay} and \cite{Saa} to the physical regime of nonminimally
coupled neutral scalar fields with $0<\xi<1/2$). Our theorem, to be
proved below, is based on simple physical restrictions imposed by
causality on the energy-momentum tensor of the fields near the
regular horizon of the black-hole spacetime.

\section{Description of the system}

We consider a non-linear physical system composed of a neutral black
hole of horizon radius $r_{\text{H}}$ and a massless scalar field
$\psi$ with nonminimal coupling to gravity. The composed
black-hole-scalar-field system is assumed to be static and
spherically symmetric, in which case the spacetime outside the
black-hole horizon is characterized by the curved line element
\cite{BekMay} (We shall use natural units in which $G=c=1$)
\begin{equation}\label{Eq1}
ds^2=-e^{\nu}dt^2+e^{\lambda}dr^2+r^2(d\theta^2+\sin^2\theta
d\phi^2)\ ,
\end{equation}
where $\nu=\nu(r)$ and $\lambda=\lambda(r)$ [Here
$(t,r,\theta,\phi)$ are the Schwarzschild coordinates]. As
explicitly proved in \cite{BekMay}, regardless of the matter content
of the curved spacetime, a non-extremal regular black hole is
characterized by the near-horizon relations \cite{Notell}
\begin{equation}\label{Eq2}
e^{-\lambda}=L\cdot x+O(x^2)\ \ \ \ \text{where}\ \ \ \ x\equiv
{{r-r_{\text{H}}}\over{r_{\text{H}}}}\ \ \ ; \ \ \ L>0\
\end{equation}
and
\begin{equation}\label{Eq3}
\lambda{'}r_{\text{H}}=-{{1}\over{x}}+O(1)\ \ \ \ ; \ \ \ \
\nu{'}r_{\text{H}}={{1}\over{x}}+O(1)\  .
\end{equation}

The curved black-hole spacetime is non-linearly and non-minimally
coupled to a real massless scalar field $\psi$ whose action is given
by \cite{BekMay}
\begin{equation}\label{Eq4}
S=S_{EH}-{1\over2}\int\big(\partial_{\alpha}\psi\partial^{\alpha}\psi+\xi
R\psi^2\big)\sqrt{-g}d^4x\ ,
\end{equation}
where the dimensionless physical parameter $\xi$ quantifies the
strength of the nonminimal coupling of the field to gravity, $R(r)$
is the scalar curvature of the spacetime, and $S_{\text{EH}}$ is the
Einstein-Hilbert action. As explicitly shown in \cite{BekMay}, in
the near-horizon $x\ll1$ region, $R$ is given by the simple
expression
\begin{equation}\label{Eq5}
R={{4L-2}\over{r^2_{\text{H}}}}\cdot[1+O(x)]\  .
\end{equation}

From the action (\ref{Eq4}) one finds the characteristic
differential equation \cite{BekMay}
\begin{equation}\label{Eq6}
\partial_{\alpha}\partial^{\alpha}\psi-\xi R\psi=0\
\end{equation}
for the nonminimally coupled scalar field. Using the metric
components (\ref{Eq1}) of the curved black-hole spacetime, one can
express the scalar radial equation in the form
\begin{equation}\label{Eq7}
\psi{''}+{1\over2}\big({{4}\over{r}}+\nu{'}-\lambda{'}\big)\psi{'}-\xi
R e^{\lambda}\psi=0\  .
\end{equation}
(Here a prime ${'}$ denotes a spatial derivative with respect to the
radial coordinate $r$).

The action (\ref{Eq4}) also yields the compact expressions
\cite{BekMay}
\begin{equation}\label{Eq8}
T^{t}_{t}=e^{-\lambda}{{\xi(4/r-\lambda{'})\psi\psi{'}+(2\xi-1/2)(\psi{'})^2+2\xi\psi\psi{''}}
\over{1-8\pi\xi\psi^2}}\
\end{equation}
and
\begin{equation}\label{Eq9}
T^{t}_{t}-T^{\phi}_{\phi}=e^{-\lambda}{{\xi(2/r-\nu{'})\psi\psi{'}}
\over{1-8\pi\xi\psi^2}}
\end{equation}
for the components of the energy-momentum tensor. As explicitly
proved in \cite{BekMay}, regardless of the matter content of the
theory, a regular hairy black-hole spacetime must be characterized
by finite mixed components of the energy-momentum tensor:
\begin{equation}\label{Eq10}
\{|T^{t}_{t}|,|T^{r}_{r}|,|T^{\theta}_{\theta}|,|T^{\phi}_{\phi}|\}<\infty\
.
\end{equation}
In addition, it was proved in \cite{BekMay} that causality
requirements enforce the characteristic inequalities
\footnotemark[1]
\begin{equation}\label{Eq11}
|T^{\theta}_{\theta}|=|T^{\phi}_{\phi}|\leq|T^{t}_{t}|\geq|T^{r}_{r}|\
\end{equation}
on the components of the energy-momentum tensors of physically
acceptable systems. Note that the relations \cite{BekMay}
\begin{equation}\label{Eq12}
\text{sgn}(T^{t}_{t})=\text{sgn}(T^{t}_{t}-T^{r}_{r})=\text{sgn}(T^{t}_{t}-T^{\phi}_{\phi})\
\end{equation}
provide necessary conditions for the validity of the characteristic
energy conditions (\ref{Eq11}).

\footnotetext[1]{As explicitly shown by Bekenstein and Mayo
\cite{BekMay}, for spherically symmetric spacetimes one can write
$\epsilon=-T^{t}_{t}-\sum_{i=1}^{3}c^2_i(T^{t}_{t}-T^{i}_{i})$ and
$j^{\mu}j_{\mu}=-(T^{t}_{t})^2-\sum_{i=1}^{3}c^2_i[(T^{t}_{t})^2-(T^{i}_{i})^2]$,
where $\epsilon\equiv T_{\mu\nu}u^{\mu}u^{\nu}$ and
$j^{\mu}\equiv-T^{\mu}_{\nu}u^{\nu}$ are respectively the energy
density and the Poynting vector according to a physical observer
with a 4-velocity $u^{\nu}$, and the coefficients
$\{c_i\}_{i=0}^{3}$ are characterized by the normalization condition
$-c^2_0+\sum_{i=1}^{3}c^2_i=-1$ (this relation guarantees that
$u^{\mu}u_{\mu}=-1$ \cite{BekMay}). For physically acceptable
systems in which the transfer of energy is not superluminal, the
energy density should be of the same sign as $-T^{t}_{t}$ and the
Poynting vector should be non-spacelike ($j^{\mu}j_{\mu}\leq0$) for
all observers \cite{BekMay} (that is, for all choices of the
coefficients $\{c_i\}_{i=0}^{3}$). These physical requirements yield
the characteristic energy conditions (\ref{Eq11}) \cite{BekMay}.}

\section{The no-hair theorem for static nonminimally coupled massless scalar fields}

In the present section we shall explicitly prove that a spherically
symmetric non-extremal neutral black hole {\it cannot} support
non-linear hair made of static nonminimally coupled massless scalar
fields.

We start our proof with the scalar radial equation (\ref{Eq7})
which, in the near-horizon $x\ll1$ region, can be written in the
form [see Eqs. (\ref{Eq2}), (\ref{Eq3}), and (\ref{Eq5})]
\begin{equation}\label{Eq13}
{{d^2\psi}\over{dx^2}}+{{1}\over{x}}{{d\psi}\over{dx}}+{{\beta}\over{x}}\psi=0\
\ \ \ ; \ \ \ \ \beta\equiv \xi(2-4L)/L\  .
%\psi{''}+{{1}\over{xr_{\text{H}}}}\psi{'}-{{\beta}\over{x}}\psi=0\ \
%\ \ ; \ \ \ \ \beta\equiv \xi R(r_{\text{H}})/L\  .
\end{equation}
The general mathematical solution of the ordinary differential
equation (\ref{Eq13}) can be expressed in terms of the familiar
Bessel functions (see Eq. 9.1.53 of \cite{Abram})
\begin{equation}\label{Eq14}
\psi(x)=A\cdot J_0(2\beta^{1/2}x^{1/2})+ B\cdot
Y_0(2\beta^{1/2}x^{1/2})\ \ \ \ \text{for}\ \ \ \ x\ll1\  ,
\end{equation}
where $\{A,B\}$ are constants. Using Eqs. 9.1.8 and 9.1.12 of
\cite{Abram}, one finds the asymptotic near-horizon behavior
\begin{equation}\label{Eq15}
\psi(x\to0)=A\cdot[1-\beta x+O(x^2)]+ B\cdot [\pi^{-1}\ln(\beta
x)+O(1)]\
\end{equation}
of the radial scalar function. Substituting Eqs. (\ref{Eq2}),
(\ref{Eq3}), and (\ref{Eq15}) into Eq. (\ref{Eq8}) and taking
cognizance of the energy condition (\ref{Eq10}) \cite{BekMay}, one
immediately realizes that the coefficient of the singular term in
the asymptotic near-horizon solution (\ref{Eq15}) should vanish
\cite{Noteb0}:
\begin{equation}\label{Eq16}
B=0\  .
\end{equation}

We therefore find that the nonminimally coupled scalar field is
characterized by the near-horizon behavior
\begin{equation}\label{Eq17}
\psi(x\ll1)=A\cdot J_0(2\beta^{1/2}x^{1/2})\  .
\end{equation}
Substituting (\ref{Eq17}) into (\ref{Eq8}) and (\ref{Eq9}) and using
the near-horizon relations (\ref{Eq2}) and (\ref{Eq3}), one obtains
the simple expressions
\begin{equation}\label{Eq18}
T^{t}_{t}=\xi\cdot{{L\psi\psi{'}}\over{r_{\text{H}}(1-8\pi\xi\psi^2)}}\cdot[1+O(x)]\
%\ \ \ \text{for}\ \ \ \ x\ll1\
\end{equation}
and
\begin{equation}\label{Eq19}
T^{t}_{t}-T^{\phi}_{\phi}=-\xi\cdot{{L\psi\psi{'}}\over{r_{\text{H}}(1-8\pi\xi\psi^2)}}\cdot[1+O(x)]\
%\ \ \ \text{for}\ \ \ \ x\ll1\
\end{equation}
for the components of the energy-momentum tensor in the near-horizon
$x\ll1$ region. We immediately find from (\ref{Eq18}) and
(\ref{Eq19}) the near-horizon relation
\begin{equation}\label{Eq20}
T^{t}_{t}=-(T^{t}_{t}-T^{\phi}_{\phi})\  ,
\end{equation}
in {\it contradiction} with the characteristic relation (\ref{Eq12})
imposed by causality on the energy-momentum components of physically
acceptable systems.

\section{Summary}

In this compact analysis, we have proved that {\it if} a spherically
symmetric neutral black hole can support non-linear configurations
made of nonminimally coupled massless scalar fields, then in the
near-horizon $(r-r_{\text{H}})/r_{\text{H}}\ll1$ region the energy
momentum components of the fields must be characterized by the
relation $T^{t}_{t}=-(T^{t}_{t}-T^{\phi}_{\phi})$ [see Eq.
(\ref{Eq20})]. However, one realizes that this near-horizon behavior
is in {\it contradiction} with the characteristic relation
$\text{sgn}(T^{t}_{t})=\text{sgn}(T^{t}_{t}-T^{\phi}_{\phi})$ [see
Eq. (\ref{Eq12})] which, as explicitly proved in \cite{BekMay}, is
imposed by causality on the energy-momentum components of generic
physically acceptable systems. Thus, there are no physically
acceptable solutions for the eigenfunction of the external
nonminimally coupled massless scalar fields except the trivial one,
$\psi\equiv0$.

We therefore conclude that spherically symmetric neutral black holes
cannot support static configurations made of nonminimally coupled
massless scalar fields with {\it generic} values of the
dimensionless physical parameter $\xi$.

%\newpage

\bigskip
\noindent
{\bf ACKNOWLEDGMENTS}
\bigskip

This research is supported by the Carmel Science Foundation. I would
like to thank Yael Oren, Arbel M. Ongo, Ayelet B. Lata, and Alona B.
Tea for helpful discussions.

%\newpage


\begin{thebibliography}{99}
\bibitem{Notell} As shown in \cite{BekMay}, the expansion coefficient
is given by $L\equiv 1+8\pi T^{t}_{t}(r_{\text{H}})
r^2_{\text{H}}>0$, where $-T^{t}_{t}(r_{\text{H}})$ is the energy
density of the matter fields at the black-hole horizon.
\end{thebibliography}
\end{document}